\documentclass[aps,prd,superscriptaddress,showpacs,preprint]{revtex4}
\usepackage{graphicx, bm}
\usepackage[usenames]{color}

\begin{document}
\draft
\title{ Model-independent sensibility studies for the anomalous dipole moments of the $\nu_\tau$ at the CLIC
        based $\gamma e^-$ colliders }

\author{A. A. Billur\footnote{abillur@cumhuriyet.edu.tr}}
\affiliation{\small Deparment of Physics, Cumhuriyet University, 58140, Sivas, Turkey.\\}

\author{M. K\"{o}ksal\footnote{mkoksal@cumhuriyet.edu.tr}}
\affiliation{\small Deparment of Optical Engineering, Cumhuriyet University, 58140, Sivas, Turkey.\\}

\author{ A. Guti\'errez-Rodr\'{\i}guez\footnote{alexgu@fisica.uaz.edu.mx}}
\affiliation{\small Facultad de F\'{\i}sica, Universidad Aut\'onoma de Zacatecas\\
         Apartado Postal C-580, 98060 Zacatecas, M\'exico.\\}

\author{ M. A. Hern\'andez-Ru\'{\i}z\footnote{mahernan@uaz.edu.mx}}
\affiliation{\small Unidad Acad\'emica de Ciencias Qu\'{\i}micas, Universidad Aut\'onoma de Zacatecas\\
         Apartado Postal C-585, 98060 Zacatecas, M\'exico.\\}

\date{\today}

\begin{abstract}

To improve the theoretical prediction of the anomalous dipole moments of the $\tau$-neutrino, we have carried out
a study through the process $\gamma e^- \to \tau \bar\nu_\tau \nu_e$, which represents an excellent and useful option in
determination of these anomalous parameters. To study the potential of the process $\gamma e^- \to \tau \bar\nu_\tau \nu_e$,
we apply a future high-energy and high-luminosity linear electron-positron collider, such as the CLIC, with
$\sqrt{s}=380, 1500, 3000\hspace{0.8mm}GeV$ and ${\cal L}=10, 50, 100, 200, 500, 1000, 1500, 2000, 3000\hspace{0.8mm}fb^{-1}$,
and we consider systematic uncertainties of $\delta_{sys}=0, 5, 10\%$. With these elements, we present a comprehensive and
detailed sensitivity study on the total cross-section of the process $\gamma e^- \to \tau \bar\nu_\tau \nu_e$, as well as on the
dipole moments $\mu_{\nu_\tau}$ and $d_{\nu_\tau}$ at the $95\%$ C.L., showing the feasibility of such process at the CLIC
at the $\gamma e^-$ mode with unpolarized and polarized electron beams.

\end{abstract}

\pacs{14.60.St, 13.40.Em\\
Keywords: Non-standard-model neutrinos, Electric and Magnetic Moments.}

\vspace{5mm}

\maketitle

\section{Introduction}

The magnetic and electric dipole moments of the neutrino ($\nu$MM) and ($\nu$EDM) are one of the most sensitive
probes of physics beyond the Standard Model (BSM). On this topic, in the original formulation of the Standard
Model (SM) \cite{Glashow,Weinberg,Salam} neutrinos are massless particles with zero $\nu$MM. However, in the
minimally extended SM containing gauge-singlet right-handed neutrinos, the $\nu$MM induced by radiative corrections
is unobservably small, $\mu_\nu=3eG_F m_{\nu_i}/(8\sqrt{2}\pi^2)\simeq 3.1\times 10^{-19}(m_{\nu_i}/1 \hspace{1mm} eV)\mu_B$,
where $\mu_B=e/2m_e$ is the Bohr magneton \cite{Fujikawa,Shrock}. Present experimental limits on these $\nu$MM are several
orders of magnitude larger, so that a MM close to these limits would indicate a window for probing effects induced
by new physics BSM \cite{Fukugita}. Similarly, a $\nu$EDM will also point to new physics and will be of relevance
in astrophysics and cosmology, as well as terrestrial neutrino experiments \cite{Cisneros}.

A fundamental challenge of the particle physics community is to determine the Majorana or Dirac nature of the neutrino. For respond
to this challenge, experimentalist are exploring different reactions where the Majorana nature may manifest \cite{Zralek}. About this
topic, the study of neutrino magnetic moments is, in principle, a way to distinguish between Dirac and Majorana neutrinos since the
Majorana neutrinos can only have flavor changing, transition magnetic moments while the Dirac neutrinos can only have flavor conserving one.

Another fundamental challenge posed by the scientific community is the following: are the laws of physics the same for matter and anti-matter,
or are matter and anti-matter intrinsically different ? It is possible that the answer to this problem may hold the key to solving the
mystery of the matter-dominated Universe ? A. Sakharov proposed a solution to this problem \cite{Sakharov}, your proposal requires
the violation of a fundamental symmetry of nature: the CP symmetry. The study of CP violation addresses this problem,
as well as many other predicted for the SM. The SM predict CP violation, which is necessary for the existence of the electric dipole moments
(EDM) of a variety physical systems. The EDM provides a direct experimental probe of CP violation \cite{Christenson,Abe,Aaij}, a feature of
the SM and beyond SM physics. The signs of new physics can be analyzed by investigating the electromagnetic dipole moments of the tau-neutrino,
such as its MM and EDM. In recent years, the $\nu_\tau$EDM have received much attention because the experimental sensitivity is expected to
improve considerably in the future. Precise measurement of the $\nu_\tau$EDM is an important probe of CP violation.

In the case of the $\nu_e$MM and $\nu_\mu$MM, the best current sensitivity limits are derived from reactor neutrino experiment
GEMMA \cite{Bed} and of the Liquid Scintillator Neutrino Detector (LSND) experiment \cite{Auerbach}, respectively. The obtained
sensitivity limits are

\begin{equation}
\mu^{exp}_{\nu_e}= 2.9 \times 10^{-11}\mu_B, \hspace{3mm} 90\% \hspace{2mm} C.L. \hspace{5mm} \mbox{[GEMMA]} \hspace{5mm} \mbox{\cite{Bed}},
\end{equation}

\begin{equation}
\mu^{exp}_{\nu_\mu}= 6.8 \times 10^{-10}\mu_B, \hspace{3mm} 90\% \hspace{2mm} C.L. \hspace{5mm} \mbox{[LSND]} \hspace{5mm} \mbox{\cite{Auerbach}},
\end{equation}

\noindent these limits are eight-nine orders of magnitude weaker than the SM prediction.

For the electric dipole moments $d_{\nu_e, \nu_\mu}$ \cite{Aguila} the best bounds are:

 \begin{equation}
d_{\nu_e, {\nu}_\mu} < 2\times 10^{-21} (e cm), \hspace{5mm} 95\%\hspace{0.8mm}C.L..\\
\end{equation}

For the tau-neutrino, the bounds on their dipole moments are less restrictive, and therefore it is worth investigating
in deeper way their electromagnetic properties. The $\tau$-neutrino correspond to the more massive third generation of
leptons and possibly possesses the largest mass and the largest magnetic and electric dipole moments.

Table I of Ref. \cite{Gutierrez10}, summary the current experimental and theoretical bounds on the anomalous dipole moments
of the tau-neutrino. The present experimental bounds on the anomalous magnetic moment of the tau-neutrino has been reported
by different experiments at Borexino \cite{Borexino}, E872 (DONUT) \cite{DONUT}, CERN-WA-066 \cite{A.M.Cooper}, and at LEP \cite{L3}.
In addition, other limits on the $\nu_\tau$MM and $\nu_\tau$EDM in different context are reported in Refs.
\cite{Gutierrez12,Gutierrez11,Gutierrez10,Gutierrez9,Gutierrez8,Data2016,Gutierrez7,Gutierrez6,Aydin,Gutierrez5,Gutierrez4,
Gutierrez3,Keiichi,Aytekin,Gutierrez2,Gutierrez1,DELPHI,Escribano,Gould,Grotch,Sahin,Sahin1}.

For the study of the dipole moments of the $\nu_\tau$ we consider the process $\gamma e^- \to \tau \bar\nu_\tau \nu_e$,
in the presence of anomalous magnetic and electric dipole couplings $\mu_{\nu_\tau}$ and $d_{\nu_\tau}$, respectively.
The set Feynman diagrams are given in Fig. 1. The final state given by $\gamma e^- \to \tau \bar\nu_\tau \nu_e$ is
considered with the subsequent decay of the tau-lepton through two different decay channels, the leptonic decay channel
and the hadronic decay channel.

The neutrino is a neutral particle, therefore its electromagnetic properties appear only at loop level. However, a method
of studying these properties on a model-independent form is to consider the effective neutrino-photon interaction. In this
regard, the most general expression consistent with Lorentz and electromagnetic gauge invariance,
for the tau-neutrino electromagnetic vertex may be parameterized in terms of four form factors \cite{Nieves,Kayser,Kayser1}:

\begin{equation}
\Gamma^{\alpha}=eF_{1}(q^{2})\gamma^{\alpha}+\frac{ie}{2m_{\nu_\tau}}F_{2}(q^{2})\sigma^{\alpha
\mu}q_{\mu}+\frac{e}{2m_{\nu_\tau}}F_3(q^2)\gamma_5\sigma^{\alpha\mu}q_\mu +eF_4(q^2)\gamma_5(\gamma^\alpha - \frac{q\llap{/}q^\alpha}{q^2}),
\end{equation}

\noindent where $e$ is the charge of the electron, $m_{\nu_\tau}$ is the mass of the tau-neutrino, $q^\mu$
is the photon momentum, and $F_{1, 2, 3, 4}(q^2)$ are the electromagnetic form factors of the neutrino,
corresponding to charge radius, MM, EDM and anapole moment (AM),
respectively, at $q^2=0$ \cite{Escribano,Vogel,Bernabeu1,Bernabeu2,Dvornikov,Giunti,Broggini}.

The future $e^+ e^-$ linear colliders are being designed to function also as $\gamma\gamma$ or $\gamma e^-$ colliders with the photon
beams generated by laser-backscattering method, in these modes the flexibility in polarizing both lepton and photon beams will allow unique
opportunities to analyze the tau-neutrino properties and interactions. It is therefore conceivable to exploit the sensitivity of these
$\gamma e^-$ colliders based on $e^+ e^-$ linear colliders of center-of-mass energies of $380-3000\hspace{0.8mm} GeV$. See Refs.
\cite{Ginzburg,Telnov,Milburn,Arutyunyan} for a detailed description of the $\gamma\gamma$ and $\gamma e^-$ colliders.

To study the sensitivity on the anomalous dipole moments of the tau-neutrino, we consider a future high-energy and high-luminosity
linear electron positron collider, such as the Compact Linear Collider (CLIC) \cite{Accomando,Dannheim,Abramowicz}, with center-of-mass energies
of $\sqrt{s}= 380, 1500, 3000 \hspace{0.8mm} GeV$ and luminosities of ${\cal L}=10, 50, 100, 200, 500, 1000, 1500, 2000,
3000\hspace{0.8mm}fb^{-1}$. Furthermore, we apply systematic uncertainties of $\delta_{sys}=0, 5, 10\%$, as well as polarized
electron beams which affect the total and angular cross-section.

This article is organized as follows. In Section II, we study the total cross-section and the dipole moments of the tau-neutrino
through the process $\gamma e^- \to \tau \bar\nu_\tau \nu_e$ with unpolarized and polarized beams. The Section III is devoted to
our conclusions.

\vspace{5mm}

\section{Cross-section of the process $\gamma e^- \to \tau \bar\nu_\tau \nu_e$  with unpolarized
and polarized beams }

\vspace{3mm}

The CLIC physics program \cite{Accomando,Dannheim,Abramowicz} is very broad and rich which complements the physics program of the LHC. Furthermore,
it provides a unique opportunity to study $\gamma\gamma$ and $\gamma e^-$ interactions with energies and luminosities similar to
those in $e^+e^-$ collisions.

On the other hand, although many particles and processes can be produced in both colliders $e^+e^-$ and $\gamma\gamma$, $\gamma e^-$
the reactions are different and will give complementary and very valuable information about new physics phenomena, such as is the
case of the dipole moments of the tau-neutrino which we study through the process $\gamma e^- \to \tau \bar\nu_\tau \nu_e$.
Fig. 1 shows the Feynman diagrams corresponding to said process. Our numerical analyses are carried out
using the CALCHEP 3.6.30 \cite{Calhep} package, which can computate the Feynman diagrams, integrate over multiparticle phase
space and event simulation.

We evaluate the total cross-section of the process $\gamma e^- \to \tau \bar\nu_\tau \nu_e$ as a function of the anomalous form
factors $F_2$, $F_3$ and $(F_2, F_3)$ and tau lepton decays hadronic and leptonic modes are considered.

In order to evaluate the total cross-section $\sigma(\gamma e^- \to \tau \bar\nu_\tau \nu_e)$ and to probe the dipole moments
$\mu_{\nu_\tau}$ and $d_{\nu_\tau}$, we examine the potential of CLIC based $\gamma e^-$ colliders with the main parameters
given in Table I. In addition, in order to suppress the backgrounds and optimize the signal sensitivity, we impose for our
study the following kinematic basic acceptance cuts for $\tau \bar\nu_\tau \nu_e$ events at the CLIC:

\begin{table}
\caption{Benchmark parameters of the CLIC based $\gamma e^-$ colliders \cite{Accomando,Dannheim,Abramowicz}.}
\label{tab:1}
\begin{tabular}{|c|c|c|}
\hline
CLIC       &  $\sqrt{s}\hspace{0.8mm}(GeV)$     &  ${\cal L}(fb^{-1})$      \\
\hline\hline
First stage & 380 & 10, 50, 100, 200, 500     \\
\hline
Second stage & 1500 & 10, 50, 100, 200, 500, 1000, 1500     \\
\hline
Third stage & 3000 & 10, 100, 500, 1000, 2000, 3000     \\
\hline
\end{tabular}
\end{table}


\begin{eqnarray}
\begin{array}{c}
\noindent \mbox{Cut-1:} \hspace{2mm} p^\nu_T > 15\hspace{0.8mm}GeV,\\
\hspace{-1cm} \mbox{Cut-2:} \hspace{2mm} \eta^\tau < 2.5,\\
\noindent \mbox{Cut-3:} \hspace{2mm} p^\tau_T > 20\hspace{0.8mm}GeV,
\end{array}
\end{eqnarray}

\noindent where in these equations $p^{\nu, \tau}_T$ is the transverse momentum of the final state particles and
$\eta^\tau$ is the pseudorapidity which reduces the contamination from other particles misidentified as tau.

Furthermore, to study the sensitivity to the parameters of the $\gamma e^- \to \tau \bar\nu_\tau \nu_e$ process
we use the chi-squared function. The $\chi^2$ function is defined as follows
\cite{Gutierrez10,Koksal0,Ozguven,Koksal1,Koksal2,Billur,Sahin1}

\begin{equation}
\chi^2=\Biggl(\frac{\sigma_{SM}-\sigma_{NP}(\sqrt{s}, \mu_{\nu_\tau}, d_{\nu_\tau})}{\sigma_{SM}\sqrt{(\delta_{st})^2
+(\delta_{sys})^2}}\Biggr)^2,
\end{equation}

\noindent where $\sigma_{NP}(\sqrt{s}, \mu_{\nu_\tau}, d_{\nu_\tau})$ is the total cross-section including
contributions from the SM and new physics, $\delta_{st}=\frac{1}{\sqrt{N_{SM}}}$ is the statistical error
and $\delta_{sys}$ is the systematic error. The number of events is given by $N_{SM}={\cal L}_{int}\times \sigma_{SM}\times BR$,
where ${\cal L}_{int}$ is the integrated CLIC luminosity. The main tau-decay branching ratios are given in Ref. \cite{Data2016}.
In addition, as the tau-lepton decays roughly $35\%$ of the time leptonically and $65\%$ of the time to one or more hadrons,
then for the signal the following cases are consider: a) only the leptonic decay channel of the tau-lepton, b) only the hadronic
decay channel of the tau-lepton.

Systematic uncertainties arise due to many factors when identifying to the tau-lepton. Tau tagging efficiencies have been studied using
the International Large Detector (ILD) \cite{ild}, a proposed detector concept for the International Linear Collider (ILC). However,
we do not have any CLIC reports \cite{7,8} to know exactly what the systematic uncertainties are for our processes, we can assume
some of their general values. Due to these difficulties, tau identification efficiencies are always calculated for specific processes,
luminosity, and kinematic parameters. These studies are currently being carried out by various groups for selected productions. For
realistic efficiency, we need a detailed study for our specific process and kinematic parameters. For all of these reasons, kinematic
cuts contain some general values chosen by lepton identification detectors and efficiency is therefore considered within systematic errors.
It may be assumed that this accelerator will be built in the coming years and the systematic uncertainties will be lower as detector technology
develops in the future.

It is also important to consider the impact of the polarization electron beam on the collider. On this, the CLIC baseline design
supposes that the electron beam can be polarized up to $\mp80\%$ \cite{Moortgat,Ari}. By choose different beam polarizations it is
possible to enhance or suppress different physical processes. Furthermore, in the study of the process $\gamma e^- \to \tau \bar\nu_\tau \nu_e$
the polarization electron beam may lead to a reduction of the measurement uncertainties, either by increasing the signal
cross-section, therefore reducing the statistical uncertainty, or by suppressing important backgrounds.

The general formula for the cross-section for arbitrary polarized $e^+ e^-$ beams is give by \cite{Moortgat}

\begin{eqnarray}
\sigma(P_{e^-},P_{e^+})=&&\frac{1}{4}[(1+P_{e^-})(1+P_{e^+})\sigma_{++}+(1-P_{e^-})(1-P_{e^+})\sigma_{--}\nonumber\\
&&+(1+P_{e^-})(1-P_{e^+})\sigma_{+-}+(1-P_{e^-})(1+P_{e^+})\sigma_{-+}],
\end{eqnarray}

\noindent where $P_{e^-} (P_{e^+})$ is the polarization degree of the electron (positron) beam, while $\sigma_{-+}$
stands for the cross-section for completely left-handed polarized $e^-$ beam $P_{e^-}=-1$ and completely right-handed
polarized $e^+$ beam $P_{e^+}=1$, and other cross-sections $\sigma_{--}$, $\sigma_{++}$ and $\sigma_{+-}$ are defined
analogously.

For $\gamma e^-$ collider, the most promising mechanism to generate energetic photon beams in a linear collider is Compton
backscattering. The photon beams are generated by the Compton backscattered of incident electron and laser beams just before
the interaction point. The total cross-sections of the process $\gamma e^- \to \tau \bar\nu_\tau \nu_e$ are

\begin{eqnarray}
\sigma=\int f_{\gamma/e}(x)d\hat{\sigma}dE_{1}.
\end{eqnarray}

\noindent In this equation, the spectrum of Compton backscattered photons  \cite{Ginzburg,Telnov} is given by

 \begin{eqnarray}
 f_{\gamma}(y)=\frac{1}{g(\zeta)}[1-y+\frac{1}{1-y}-
 \frac{4y}{\zeta(1-y)}+\frac{4y^{2}}{\zeta^{2}(1-y)^{2}}] ,
 \end{eqnarray}

\noindent where

 \begin{eqnarray}
 g(\zeta)=(1-\frac{4}{\zeta}-\frac{8}{\zeta^2})\log{(\zeta+1)}+
 \frac{1}{2}+\frac{8}{\zeta}-\frac{1}{2(\zeta+1)^2} ,
 \end{eqnarray}

\noindent with

 \begin{eqnarray}
 y=\frac{E_{\gamma}}{E_{e}} , \;\;\;\; \zeta=\frac{4E_{0}E_{e}}{M_{e}^2}
 ,\;\;\;\; y_{max}=\frac{\zeta}{1+\zeta}.
 \end{eqnarray}

\noindent Here, $E_{0}$ and $E_{e}$ are  energy of the incoming laser photon and initial energy of the electron beam before
Compton backscattering and $E_{\gamma}$ is the energy of the backscattered photon. The maximum value of $y$ reaches 0.83
when $\zeta=4.8$.

\subsection{Cross-section of the process $\gamma e^- \to \tau \bar\nu_\tau \nu_e$ and dipole moments of the
$\nu_\tau$ with unpolarized electrons beams}

As the first observable, we consider the total cross-section. Figs. 2 and 3 summarize the total cross-section of the process
$\gamma e^- \to \tau \bar\nu_\tau \nu_e$ with unpolarized electrons beams and as a function of the anomalous couplings $F_2 (F_3)$.
We use the three stages of the center-of-mass energy of the CLIC given in Table I. The total cross-section clearly shows a
strong dependence with respect to the anomalous parameters $F_2$, $F_3$, as well as with the center-of-mass energy of the collider
$\sqrt{s}$.

The total cross-section of the process $\gamma e^- \to \tau \bar\nu_\tau \nu_e$ as a function of $F_2$ and $F_3$ with the benchmark
parameters of the CLIC given in Table I is shown in Figs. 4-6. The total cross-section increases with the increase in the center-of-mass
energy of the collider and strongly depends on anomalous couplings $F_2$ and $F_3$.

In order to investigate the signal more comprehensively, we show the bounds contours depending on integrated luminosity at the $95\%$ C.L.
on the $(F_2, F_3)$ plane for $\sqrt{s}=380, 1500, 3000 \hspace{0.8mm}GeV$ in Figs. 7-9. At $95\%$ C.L. and $\sqrt{s}= 3000 \hspace{0.8mm}GeV$,
we can see that the correlation region of $F_2\hspace{1mm}\in\hspace{1mm}[-2.5; 2.5]\times 10^{-5}$ and $F_3\hspace{1mm} \in\hspace{1mm}
[-2.5; 2.5]\times 10^{-5}$ can be excluded with integrate luminosity ${\cal L}=100\hspace{0.8mm}fb^{-1}$. If the integrated luminosity
is increased to ${\cal L}=3000\hspace{0.8mm}fb^{-1}$, the excluded region will expand into $F_2\hspace{1mm}\in \hspace{1mm} [-1, 1]\times 10^{-5}$
and $F_3\hspace{1mm}\in \hspace{1mm} [-1, 1]\times 10^{-5}$.

\subsection{Cross-section of the process $\gamma e^- \to \tau \bar\nu_\tau \nu_e$ and dipole moments of the
$\nu_\tau$ with polarized electrons beams}

We consider the total cross-section of the process $\gamma e^- \to \tau \bar\nu_\tau \nu_e$ as a function of the anomalous
form factors $F_2 (F_3)$ and we perform our analysis for the CLIC running at center-of-mass energies and luminosities given in
Table I. Furthermore, in our analysis we consider the baseline expectation of an $80\%$ left-polarized electron beam.
As expected, the polarization hugely improves the total cross-section as is shown in Figs. 10 and 11. The total cross-section
is increased from about $\sigma=8\times 10^3\hspace{0.8mm} pb$ with unpolarized electron beam (see Figs. 2 and 3) to about
$\sigma=1.5\times 10^4\hspace{0.8mm} pb$ with polarized electron beam (see Figs. 10 and 11), respectively, enhancing the
statistic. The increase of the total cross-section of the process $\gamma e^- \to \tau \bar\nu_\tau \nu_e$ for the polarized
case is approximately the double of the unpolarized case. The Feynman diagram 4 of Fig. 1, gives the maximum contribution to
the total cross-section. For $P_{e^-} =-80\%$ case, this contribution is dominant due to the structure of the $We^-\nu_{e^-}$
vertex. The advantage of beam polarization is evident when compared to the corresponding unpolarized case.

In Figs. 12 and 13 we plot the $\chi^2$ versus $F_2 (F_3)$ with unpolarized $P_{e^-}=0\%$ and polarized $P_{e^-}=-80\%$
electron beam and $95\% \hspace{0.8mm}C.L.$. We plot the curves for each case, for which we have divided the interval of
$F_2 (F_3)$ into several bins. From these figures we can see that the effect of the polarized beam is to reduce the interval
of definition of $F_2\hspace{1mm}\in \hspace{1mm} [-2; 2]\times 10^{-5}$ (unpolarized case) to $F_2\hspace{1mm}\in \hspace{1mm} [-1.65; 1.65]\times 10^{-5}$
(polarized case) and $F_3\hspace{1mm}\in \hspace{1mm} [-2; 2]\times 10^{-5}$ (unpolarized case) to
$F_3\hspace{1mm}\in \hspace{1mm} [-1.65; 1.65]\times 10^{-5}$ (polarized case), respectively.

Another important observable is the transverse momentum $p^\tau_T$ of the tau lepton, the pseudorapidity $\eta^\tau$
is also important, these quantities are shown in Figs. 14 and 15. In both cases, the tau-lepton pseudorapidity and the transverse
momentum are for the SM, SM-polarized beam, $F_2$ and $F_2$-polarized beam. From Fig. 14, the $d\sigma/d\eta$ clearly shows a strong
dependence with respect to the pseudorapidity, as well as with the form factors $F_2$ and $F_2$-polarized beam. In the case of Fig. 15,
the distribution $d\sigma/dp_T ( pb/GeV)$ decreases with the increase of $p_T$ for the SM and the SM-polarized beam, while for $F_2$ and
$F_2$-polarized beam have the opposite effect. These distributions clearly show great sensitivity with respect to the anomalous form factor
$F_2$ for the cases with unpolarized and polarized electron beam. The analysis of these distributions is important to be able to discriminate
the basic acceptance cuts for $\tau \bar\nu_\tau \nu_e$ events at the CLIC.

\subsection{$90\%$ C.L. and $95\%$ C.L. bounds on the anomalous $\nu_\tau$MM and $\nu_\tau$EDM with unpolarized and polarized electron beam}

In the following we will refer to the anomalous $\nu_\tau$MM and $\nu_\tau$EDM. From Feynman diagrams for the process
$\gamma e^- \to \tau \bar\nu_\tau \nu_e$ given in Fig. 1, for the estimation of the sensitivty on the anomalous dipole moments,
we consider the following scenarios: a) unpolarized electrons beams $P_{e^-}$=0\%
and we considered only the leptonic decay channel of the tau-lepton. b) polarized electrons beams $P_{e^-}$=-80\%, and we considered
only the leptonic decay channel of the tau-lepton. c) unpolarized electrons beams $P_{e^-}$=0\% and we considered only the hadronic
decay channel of the tau-lepton. d) polarized electrons beams $P_{e^-}$=-80\% and we considered only the hadronic decay
channel of the tau-lepton. For all these scenarios, we consider the energies and luminosities for the future CLIC summarized in
Table I. In addition, we imposing kinematic cuts on $p^\nu_T$, $p^\tau_T$ and $\eta^\tau$ to suppress the backgrounds
and to optimize the signal sensitivity (see Eq. (5)), we also consider the systematic uncertainties $\delta_{sys} = 0, 5, 10 \%$.
The achievable precision in the determination of the sensibility on $\mu_{\nu_\tau}$ and the $d_{\nu_\tau}$ is summarized in
Tables II-IX.

The best sensitivity achieved for the anomalous $\mu_{\nu_\tau}$ and the $d_{\nu_\tau}$ for the case of $P_{e^-} = 0\%$,
and considering only the leptonic decay channel of the tau-lepton are $|\mu_{\nu_\tau}(\mu_B)|= 3.649\times 10^{-7}$ and
$|d_{\nu_\tau}(e cm)|=7.072\times 10^{-18}$. In the case of $P_{e^-} = -80\%$, and considering only the leptonic decay channel
of the tau-lepton the sensitivity estimates are $|\mu_{\nu_\tau}(\mu_B)|= 3.152\times 10^{-7}$ and
$|d_{\nu_\tau}(e cm)|=6.108\times 10^{-18}$. In both cases the obtained sensitivity are for the values of
$\sqrt{s}=3000\hspace{0.8mm}GeV$, ${\cal L}=3000\hspace{0.8mm} fb^{-1}$ and $95\%$ C.L. Comparing both
cases, unpolarized and polarized electron beams, we conclude that the case with polarized beams $P_{e^-} = -80\%$
improves the sensitivity on the anomalous dipole moments in $13.63\%$ with respect to the unpolarized case.

When only the hadronic decay channel of the tau-lepton is considered, the sensibility on the dipole moments is
$|\mu_{\nu_\tau}(\mu_B)|= 3.127\times 10^{-7}$, $|d_{\nu_\tau}(e cm)|=6.059\times 10^{-18}$ with $P_{e^-} = 0\%$
and $|\mu_{\nu_\tau}(\mu_B)|= 2.700\times 10^{-7}$, $|d_{\nu_\tau}(e cm)|=5.232\times 10^{-18}$ with $P_{e^-} =-80\%$,
respectively. The obtained results are with $\sqrt{s}=3000\hspace{0.8mm}GeV$, ${\cal L}=3000\hspace{0.8mm} fb^{-1}$
and $95\%$ C.L. The comparison of both cases shows that the case with polarized electron beams improves the
sensitivity of the anomalous dipole moments of the $\tau$-neutrino of $13.65\%$, with respect to the case with $P_{e^-} = 0\%$.

If now we compare the cases with leptonic decay channel and hadronic decay channel with $P_{e^-} = 0\%$ and
$\sqrt{s}=3000\hspace{0.8mm}GeV$, ${\cal L}=3000\hspace{0.8mm} fb^{-1}$ and $95\%$ C.L. the improvement in
sensitivity is $14.31\%$ for the hadronic decay channel with respect to the leptonic decay channel. Whereas for
$P_{e^-} = -80\%$ the improvement in the sensitivity is of $14.34\%$ with respect to the case of the leptonic
decay channel. These differences are expected for the different cases because the tau-lepton decays roughly
$35\%$ of the time leptonically and $65\%$ of the time hadronically.

\section{Conclusions}

We have studied the $\nu_\tau$MM and $\nu_\tau$EDM in a model-independent way. For the two options that we
considered in this paper: polarized and unpolarized electron beams, our results are sensitive to the parameters of
the collider such as the center-of-mass energy and the luminosity. Furthermore, our results are also sensitive to
the kinematic basic acceptance cuts of the final states particles $p^\nu_T$, $\eta^\tau$ and $p^\tau_T$, as well as the systematic
uncertainties $\delta_{sys}$. A good knowledge of the kinematic cuts is needed not only to improve sensitivity analyses,
but because can help to understand which are the most appropriate processes to probing in the future high-energy and
high-luminosity linear colliders, such as the CLIC. CLIC, as well as any $\gamma e^-$ Compton backscattering experiment,
offers a good laboratory to study the total cross-section and the dipole moments of the tau-neutrino through the process
$\gamma e^- \to \tau \bar\nu_\tau \nu_e$ with unpolarized and polarized electron beams.

Despite the large number of study performed in recent years on the electromagnetic properties of the tau-neutrino,
more studies are still needed to deeply understand and explain experimental observations and their comparison with
models predictions. Current and future data of the ATLAS \cite{ATLAS} and CMS \cite{V,V1} Collaborations,
as well as new analysis of already existing data sets, could help to improve our knowledge on the $\nu_\tau$MM and
$\nu_\tau$EDM \cite{Gutierrez10}.

A precision machine like CLIC is expected to help in the precise estimates of the anomalous couplings. In this paper, the process
$\gamma e^- \to \tau \bar\nu_\tau \nu_e$, which contains the neutrino to photon coupling, namely $\nu_\tau\bar\nu_\tau\gamma$
is considered. The reach of the CLIC with maximum $\sqrt{s}=3000\hspace{0.8mm}GeV$ and ${\cal L}=3000\hspace{0.8mm}fb^{-1}$
to probing the relevant observable of the process is presented. The influence of the anomalous couplings, of the kinematic cuts,
of the uncertainties systematic, as well as the polarized electron beam on the cross-section, the tau-lepton pseudorapidity
distribution and the tau-lepton transverse momentum distribution are studied. Furthermore, we estimates the sensitivity on
the anomalous $\nu_\tau$MM and $\nu_\tau$EDM. Our results are summarized in Figs. 2-15 as well as in Tables II-IX, respectively.

From our set of Figures and Tables it is evident that a suitably chosen beam polarization is found to be advantageous as illustrated
with an $80\%$ left-polarization electron beam (see Figs. 10-15). The most optimistic scenario about the sensitivity in the anomalous
dipole moments of the tau-neutrino (see Tables V and IX), yields the following results: $|\mu_{\nu_\tau}(\mu_B)|=2.998\times 10^{-7}$ and
$|d_{\nu_\tau}(e cm)|=5.598\times10^{-18}$ with $P_{e^-}=-80\%$ and we considered only the leptonic decay channel of the tau-lepton. $|\mu_{\nu_\tau}(\mu_B)|=2.475\times 10^{-7}$ and $|d_{\nu_\tau}(e cm)|=4.796\times 10^{-18}$, with $P_{e^-}=-80\%$ and we taken in
account only the hadronic decay channel of the tau-lepton. Our results show the potential and the feasibility of the process
$\gamma e^- \to \tau \bar\nu_\tau \nu_e$ at the CLIC at the $\gamma e^-$ mode.

\vspace{2cm}

\begin{center}
{\bf Acknowledgements}
\end{center}

A. G. R. and M. A. H. R acknowledges support from SNI and PROFOCIE (M\'exico).

\vspace{2cm}

\begin{table}[!ht]
\caption{Limits on the $\mu_{\nu_\tau}$ magnetic moment and $d_{\nu_\tau}$ electric dipole moment via the process $\gamma e^- \to \tau \bar\nu_\tau \nu_e$
($\gamma$ is the Compton backscattering photon) for $P_{e^-}=0\%$, and we considered only the leptonic decay channel of the tau-lepton.}
\label{tab1}
\begin{center}
\begin{tabular}{ccccc}
\hline\hline
\hspace{4cm}$90\%\hspace{0.8mm}C.L.$  &  \hspace{4cm}   $\sqrt{s}=1.5\hspace{0.8mm}TeV$\\
\hline
${\cal L}\hspace{0.8mm}(fb^{-1})$  &  $\vert \mu_{\nu_{\tau}}\vert$ ($10^{-7}$) & $\vert d_{\nu_{\tau}} \vert $($10^{-17}$)(e\,cm)   \\
\hline
 $100$       &$16.810$         &$3.258 $ \\
 $200$       &$14.130$         &$2.739$  \\
 $500$       &$11.240$         &$2.178 $  \\
 $1000$      &$9.455$          &$1.832$ \\
 $1500$      &$8.544$          &$1.655$ \\
 \hline
 $\delta_{sys}=5\%$        &$6.860\times 10^{-6}$       &$1.322\times10^{-16}$ \\
 $\delta_{sys}=10\%$       &$9.701 \times 10^{-6}$      &$1.879\times10^{-16}$  \\
\hline
\hline
\hspace{4cm} $95\%\hspace{0.8mm} C.L.$   &  \hspace{4cm}  $\sqrt{s}=1.5\hspace{0.8mm}TeV$\\
\hline
 $100$         & $18.340$         &$ 3.554$ \\
 $200$         & $15.420$         &$ 2.989$  \\
 $500$         & $12.260$         &$ 2.377$  \\
 $1000$        & $10.310$         &$ 1.999$ \\
 $1500$        & $9.3210$         &$ 1.806$ \\
  \hline
 $\delta_{sys}=5\%$         &$7.484\times 10^{-6}$        &$ 1.450 \times10^{-16}$  \\
 $\delta_{sys}=10\%$        &$10.580\times 10^{-6}$       &$ 2.051 \times10^{-16}$  \\
 \hline\hline
\end{tabular}
\end{center}
\end{table}

\begin{table}[!ht]
\caption{Limits on the $\mu_{\nu_\tau}$ magnetic moment and $d_{\nu_\tau}$ electric dipole moment via the process $\gamma e^- \to \tau \bar\nu_\tau \nu_e$
($\gamma$ is the Compton backscattering photon) for $P_{e^-}=0\%$, and we considered only the leptonic decay channel of the tau-lepton.}
\label{tab1}
\begin{center}
\begin{tabular}{ccccc}
\hline\hline
\hspace{4cm}$90\%\hspace{0.8mm}C.L.$  &  \hspace{4cm}   $\sqrt{s}=3\hspace{0.8mm}TeV$\\
\hline
${\cal L}\hspace{0.8mm}(fb^{-1})$  &  $\vert \mu_{\nu_{\tau}}\vert$ ($10^{-7}$) & $\vert d_{\nu_{\tau}} \vert $($10^{-18}$)(e\,cm)   \\
\hline
 $100$        &$7.826$          &$15.160 $ \\
 $500$        &$5.234$          &$10.140$  \\
 $1000$       &$4.402$          &$8.530 $  \\
 $2000$       &$3.702$          &$7.174$ \\
 $3000$       &$3.345$          &$6.483$ \\
 \hline
 $\delta_{sys}=5\%$        &$3.029\times 10^{-6}$       &$5.871\times10^{-17}$ \\
 $\delta_{sys}=10\%$       &$4.248 \times 10^{-6}$      &$8.233\times10^{-17}$  \\
\hline
\hline
\hspace{4cm} $95\%\hspace{0.8mm} C.L.$   &  \hspace{4cm}  $\sqrt{s}=3\hspace{0.8mm}TeV$\\
\hline
 $100$          & $8.538$         &$ 16.540$ \\
 $500$          & $5.710$         &$ 11.060$  \\
 $1000$         & $4.802$         &$ 9.306$  \\
 $2000$         & $4.038$         &$ 7.826$ \\
 $3000$         & $3.649$         &$ 7.072$ \\
  \hline
 $\delta_{sys}=5\%$         &$3.268\times 10^{-6}$        &$ 6.333 \times10^{-17}$  \\
 $\delta_{sys}=10\%$        &$4.622\times 10^{-6}$        &$ 8.957 \times10^{-17}$  \\
 \hline\hline
\end{tabular}
\end{center}
\end{table}

\begin{table}[!ht]
\caption{Limits on the $\mu_{\nu_\tau}$ magnetic moment and $d_{\nu_\tau}$ electric dipole moment via the process $\gamma e^- \to \tau \bar\nu_\tau \nu_e$
($\gamma$ is the Compton backscattering photon) for $P_{e^-}=-80\%$, and we considered only the leptonic decay channel of the tau-lepton.}
\label{tab1}
\begin{center}
\begin{tabular}{ccccc}
\hline\hline
\hspace{4cm}$90\%\hspace{0.8mm}C.L.$  &  \hspace{4cm}   $\sqrt{s}=1.5\hspace{0.8mm}TeV$\\
\hline
${\cal L}\hspace{0.8mm}(fb^{-1})$  &  $\vert \mu_{\nu_{\tau}}\vert$ ($10^{-7}$) & $\vert d_{\nu_{\tau}} \vert $($10^{-17}$)(e\,cm)   \\
\hline
 $100$        &$14.520$          &$2.814$ \\
 $200$        &$12.210$          &$2.366$  \\
 $500$        &$9.713$          &$1.882$  \\
 $1000$       &$8.167$          &$1.582$ \\
 $1500$       &$7.380$          &$1.430$ \\
 \hline
 $\delta_{sys}=5\%$        &$6.865\times 10^{-6}$       &$1.330\times10^{-16}$ \\
 $\delta_{sys}=10\%$       &$9.709 \times 10^{-6}$      &$1.881\times10^{-16}$  \\
\hline
\hline
\hspace{4cm} $95\%\hspace{0.8mm} C.L.$   &  \hspace{4cm}  $\sqrt{s}=1.5\hspace{0.8mm}TeV$\\
\hline
 $100$          & $15.840$         &$ 3.070$ \\
 $200$          & $13.320$         &$ 2.582$  \\
 $500$          & $10.590$         &$ 2.053$  \\
 $1000$         & $8.911$          &$ 1.726$ \\
 $1500$         & $8.052$          &$ 1.560$ \\
  \hline
 $\delta_{sys}=5\%$         &$7.491\times 10^{-6}$        &$ 1.451 \times10^{-16}$  \\
 $\delta_{sys}=10\%$        &$10.590\times 10^{-6}$       &$ 2.052 \times10^{-16}$  \\
 \hline\hline
\end{tabular}
\end{center}
\end{table}

\begin{table}[!ht]
\caption{Limits on the $\mu_{\nu_\tau}$ magnetic moment and $d_{\nu_\tau}$ electric dipole moment via the process $\gamma e^- \to \tau \bar\nu_\tau \nu_e$
($\gamma$ is the Compton backscattering photon) for $P_{e^-}=-80\%$, and we considered only the leptonic decay channel of the tau-lepton.}
\label{tab1}
\begin{center}
\begin{tabular}{ccccc}
\hline\hline
\hspace{4cm}$90\%\hspace{0.8mm}C.L.$  &  \hspace{4cm}   $\sqrt{s}=3\hspace{0.8mm}TeV$\\
\hline
${\cal L}\hspace{0.8mm}(fb^{-1})$  &  $\vert \mu_{\nu_{\tau}}\vert$ ($10^{-7}$) & $\vert d_{\nu_{\tau}} \vert $($10^{-18}$)(e\,cm)   \\
\hline
 $100$        &$6.762$          &$13.100 $ \\
 $500$        &$4.522$          &$8.762$  \\
 $1000$       &$3.802$          &$7.368 $  \\
 $2000$       &$3.197$          &$6.196$ \\
 $3000$       &$2.889$          &$5.598$ \\
 \hline
 $\delta_{sys}=5\%$        &$3.029\times 10^{-6}$       &$5.851\times10^{-17}$ \\
 $\delta_{sys}=10\%$       &$4.248 \times 10^{-6}$      &$8.236\times10^{-17}$  \\
\hline
\hline
\hspace{4cm} $95\%\hspace{0.8mm} C.L.$   &  \hspace{4cm}  $\sqrt{s}=3\hspace{0.8mm}TeV$\\
\hline
 $100$          & $7.377$         &$ 14.290$ \\
 $500$          & $4.933$         &$ 9.560$  \\
 $1000$         & $4.148$         &$ 8.039$  \\
 $2000$         & $3.488$         &$ 6.760$ \\
 $3000$         & $3.152$         &$ 6.108$ \\
  \hline
 $\delta_{sys}=5\%$         &$3.273\times 10^{-6}$        &$ 6.343 \times10^{-17}$  \\
 $\delta_{sys}=10\%$        &$4.629\times 10^{-6}$        &$ 8.971 \times10^{-17}$  \\
 \hline\hline
\end{tabular}
\end{center}
\end{table}

\begin{table}[!ht]
\caption{Limits on the $\mu_{\nu_\tau}$ magnetic moment and $d_{\nu_\tau}$ electric dipole moment via the process $\gamma e^- \to \tau \bar\nu_\tau \nu_e$
($\gamma$ is the Compton backscattering photon) for $P_{e^-}=0\%$, and we considered only the hadronic decay channel of the tau-lepton.}
\label{tab1}
\begin{center}
\begin{tabular}{ccccc}
\hline\hline
\hspace{4cm}$90\%\hspace{0.8mm}C.L.$  &  \hspace{4cm}   $\sqrt{s}=1.5\hspace{0.8mm}TeV$\\
\hline
${\cal L}\hspace{0.8mm}(fb^{-1})$  &  $\vert \mu_{\nu_{\tau}}\vert$ ($10^{-7}$) & $\vert d_{\nu_{\tau}} \vert $($10^{-17}$)(e\,cm)   \\
\hline
 $100$       &$14.400$         &$2.791 $ \\
 $200$       &$12.110$         &$2.347$  \\
 $500$       &$9.632$          &$1.866 $  \\
 $1000$      &$8.098$          &$1.569$ \\
 $1500$      &$7.319$          &$1.418$ \\
 \hline
 $\delta_{sys}=5\%$        &$6.869\times 10^{-6}$       &$1.329\times10^{-16}$ \\
 $\delta_{sys}=10\%$       &$9.701 \times 10^{-6}$      &$1.879\times10^{-16}$  \\
\hline
\hline
\hspace{4cm} $95\%\hspace{0.8mm} C.L.$   &  \hspace{4cm}  $\sqrt{s}=1.5\hspace{0.8mm}TeV$\\
\hline
 $100$         & $15.710$         &$ 3.045$ \\
 $200$         & $13.210$         &$ 2.560$  \\
 $500$         & $10.500$         &$ 2.036$  \\
 $1000$        & $8.837$         &$ 1.712$ \\
 $1500$        & $7.985$         &$ 1.547$ \\
  \hline
 $\delta_{sys}=5\%$         &$7.484\times 10^{-6}$        &$ 1.450 \times10^{-16}$  \\
 $\delta_{sys}=10\%$        &$10.580\times 10^{-6}$       &$ 2.051 \times10^{-16}$  \\
 \hline\hline
\end{tabular}
\end{center}
\end{table}

\begin{table}[!ht]
\caption{Limits on the $\mu_{\nu_\tau}$ magnetic moment and $d_{\nu_\tau}$ electric dipole moment via the process $\gamma e^- \to \tau \bar\nu_\tau \nu_e$
($\gamma$ is the Compton backscattering photon) for $P_{e^-}=0\%$, and we considered only the hadronic decay channel of the tau-lepton.}
\label{tab1}
\begin{center}
\begin{tabular}{ccccc}
\hline\hline
\hspace{4cm}$90\%\hspace{0.8mm}C.L.$  &  \hspace{4cm}   $\sqrt{s}=3\hspace{0.8mm}TeV$\\
\hline
${\cal L}\hspace{0.8mm}(fb^{-1})$  &  $\vert \mu_{\nu_{\tau}}\vert$ ($10^{-7}$) & $\vert d_{\nu_{\tau}} \vert $($10^{-18}$)(e\,cm)   \\
\hline
 $100$        &$6.704$          &$12.990 $ \\
 $500$        &$4.484$          &$8.689$  \\
 $1000$       &$3.771$          &$7.308 $  \\
 $2000$       &$3.171$          &$6.146$ \\
 $3000$       &$2.866$          &$5.554$ \\
 \hline
 $\delta_{sys}=5\%$        &$2.995\times 10^{-6}$       &$5.805\times10^{-17}$ \\
 $\delta_{sys}=10\%$       &$4.236 \times 10^{-6}$      &$8.209\times10^{-17}$  \\
\hline
\hline
\hspace{4cm} $95\%\hspace{0.8mm} C.L.$   &  \hspace{4cm}  $\sqrt{s}=3\hspace{0.8mm}TeV$\\
\hline
 $100$          & $7.314$         &$ 14.170$ \\
 $500$          & $4.892$         &$ 9.480$  \\
 $1000$         & $4.114$         &$ 7.972$  \\
 $2000$         & $3.460$         &$ 6.705$ \\
 $3000$         & $3.127$         &$ 6.059$ \\
  \hline
 $\delta_{sys}=5\%$         &$3.268\times 10^{-6}$        &$ 6.333 \times10^{-17}$  \\
 $\delta_{sys}=10\%$        &$4.622\times 10^{-6}$        &$ 8.956 \times10^{-17}$  \\
 \hline\hline
\end{tabular}
\end{center}
\end{table}

\begin{table}[!ht]
\caption{Limits on the $\mu_{\nu_\tau}$ magnetic moment and $d_{\nu_\tau}$ electric dipole moment via the process $\gamma e^- \to \tau \bar\nu_\tau \nu_e$
($\gamma$ is the Compton backscattering photon) for $P_{e^-}=-80\%$, and we considered only the hadronic decay channel of the tau-lepton.}
\label{tab1}
\begin{center}
\begin{tabular}{ccccc}
\hline\hline
\hspace{4cm}$90\% C.L.$  &  \hspace{4cm}   $\sqrt{s}=1.5\hspace{0.8mm}TeV$\\
\hline
 Luminosity ($fb^{-1}$)&  $\vert \mu_{\nu_{\tau}}\vert$ ($10^{-7}$) & $\vert d_{\nu_{\tau}} \vert $($10^{-17}$)(e\,cm)   \\
\hline
 $100$      & $12.440$      &$2.411 $ \\
 $200$      & $10.460$      &$2.027$  \\
 $500$      & $8.320$       &$1.612 $  \\
 $1000$     & $6.996$      &$1.355$ \\
 $1500$     & $6.322$      &$1.225$ \\
 \hline
 $\delta_{sys}=5\%$      &$6.865\times 10^{-6}$     &$1.330\times10^{-16}$ \\
 $\delta_{sys}=10\%$     &$9.709\times 10^{-6}$     &$1.881\times10^{-16}$  \\
\hline
\hline
\hspace{4cm} $95\% C.L.$   &  \hspace{4cm}  $\sqrt{s}=1.5\hspace{0.8mm}TeV$\\
\hline
$100$           & $13.570$       &$ 2.630$ \\
 $200$          & $11.410$       &$ 2.212$  \\
 $500$          & $9.078$        &$ 1.759$  \\
 $1000$         & $7.633$        &$ 1.479 $ \\
 $1500$         & $6.897$        &$1.336$ \\
  \hline
 $\delta_{sys}=5\%$      &$7.490\times 10^{-6}$       &$1.451 \times10^{-16}$  \\
 $\delta_{sys}=10\%$     &$10.590\times 10^{-6}$      &$2.052 \times10^{-16}$  \\
 \hline\hline
\end{tabular}
\end{center}
\end{table}

\begin{table}[!ht]
\caption{Limits on the $\mu_{\nu_\tau}$ magnetic moment and $d_{\nu_\tau}$ electric dipole moment via the process $\gamma e^- \to \tau \bar\nu_\tau \nu_e$
($\gamma$ is the Compton backscattering photon) for $P_{e^-}=-80\%$, and we considered only the hadronic decay channel of the tau-lepton.}
\label{tab1}
\begin{center}
\begin{tabular}{ccccc}
\hline\hline
\hspace{4cm}$90\% C.L.$  &  \hspace{4cm}   $\sqrt{s}=3\hspace{0.8mm}TeV$\\
\hline
 Luminosity ($fb^{-1}$)&  $\vert \mu_{\nu_{\tau}}\vert$ ($10^{-7}$) & $\vert d_{\nu_{\tau}} \vert $($10^{-18}$)(e\,cm)   \\
\hline
 $100$                & $5.792$            &$11.22 $ \\
 $500$                & $3.873$            &$ 7.506$  \\
 $1000$               & $3.257$            &$6.312 $  \\
 $2000$               & $2.739$            &$5.307$ \\
 $3000$               & $2.475$            &$4.976$ \\
 \hline
 $\delta_{sys}=5\%$&$3.000\times 10^{-6}$ &$5.814\times10^{-17}$ \\
 $\delta_{sys}=10\%$&$4.243 \times 10^{-6}$&$8.222\times10^{-17}$  \\
\hline
\hline
\hspace{4cm} $95\% C.L.$   &  \hspace{4cm}  $\sqrt{s}=3\hspace{0.8mm}TeV$\\
\hline
$100$                & $6.319$          &$ 12.240$ \\
 $500$               & $4.226$          &$ 8.189$  \\
 $1000$              & $3.553$          &$ 6.886$  \\
 $2000$              & $2.988$          &$ 5.791 $ \\
 $3000$              & $2.700$          & $5.232$ \\
  \hline
 $\delta_{sys}=5\%$&$3.273\times 10^{-6}$&$6.343 \times10^{-17}$  \\
 $\delta_{sys}=10\%$&$4.629\times 10^{-6}$&$8.971 \times10^{-17}$  \\
 \hline\hline
\end{tabular}
\end{center}
\end{table}

\newpage

\begin{figure}[t]
\centerline{\scalebox{0.65}{\includegraphics{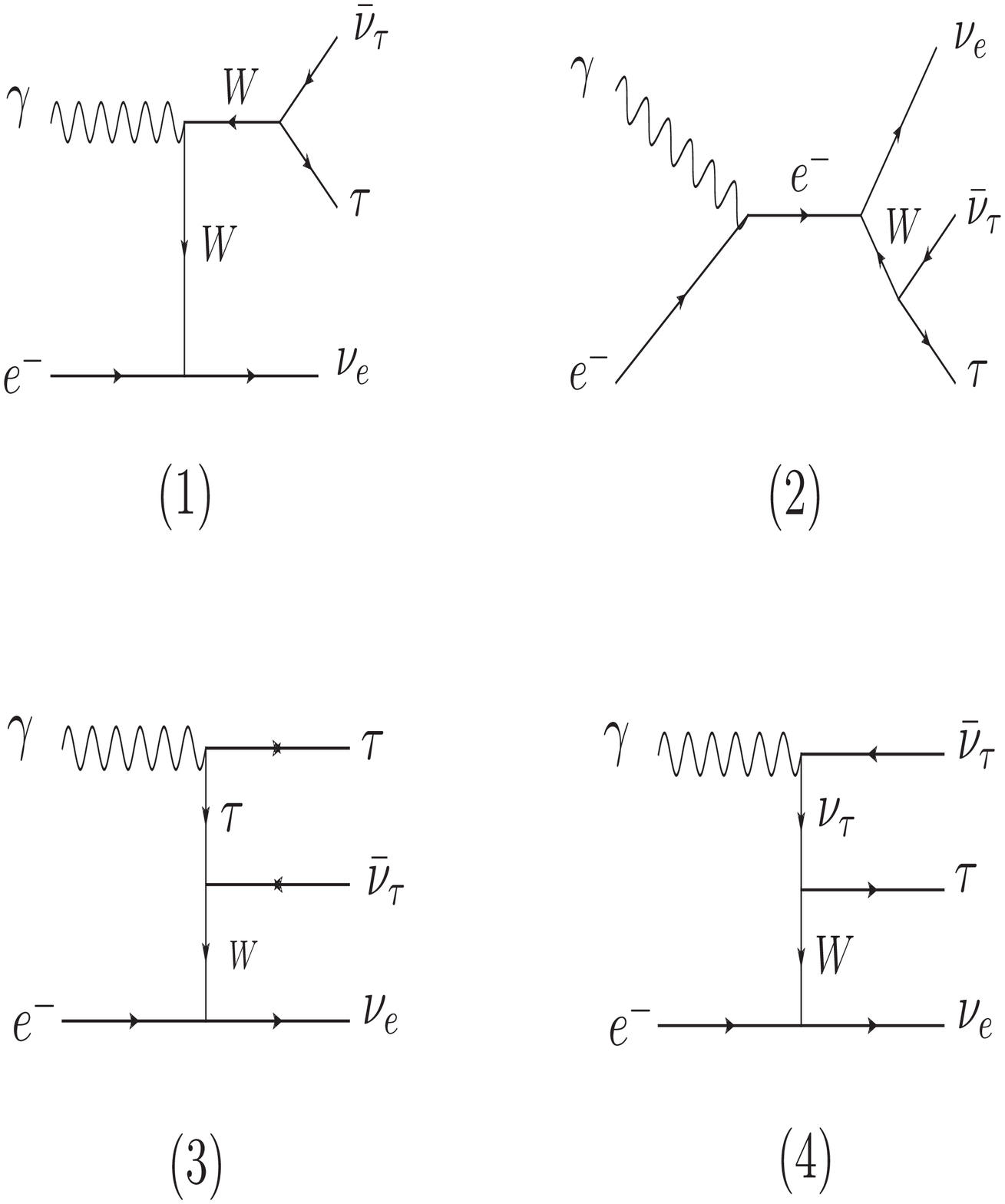}}}
\caption{ \label{fig:gamma} The Feynman diagrams contributing to the process
$\gamma e^- \to \tau \bar\nu_\tau \nu_e $.}
\end{figure}

\begin{figure}[t]
\centerline{\scalebox{1.2}{\includegraphics{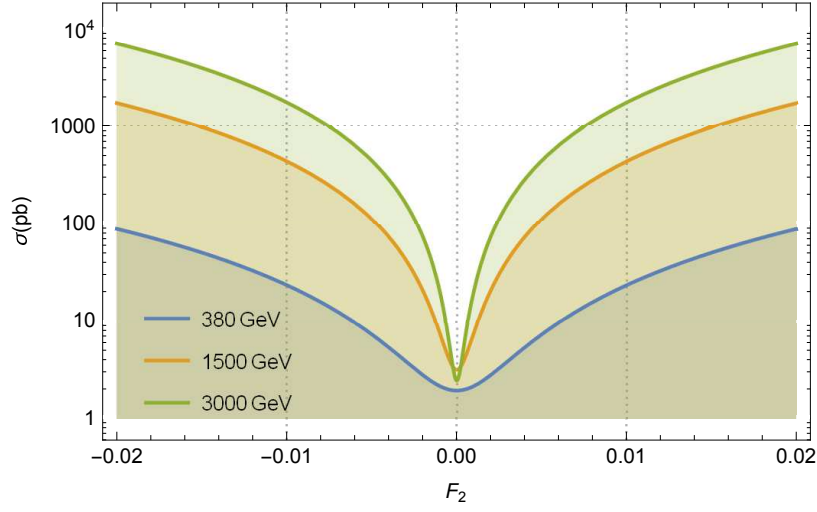}}}
\caption{ \label{fig:gamma1} The total cross-section of the process $\gamma e^- \to \tau \bar\nu_\tau \nu_e $
as a function of the anomalous coupling $F_2$ for three different center-of-mass energies $\sqrt{s}=380, 1500, 3000\hspace{0.8mm}GeV$.}
\end{figure}

\begin{figure}[t]
\centerline{\scalebox{1.2}{\includegraphics{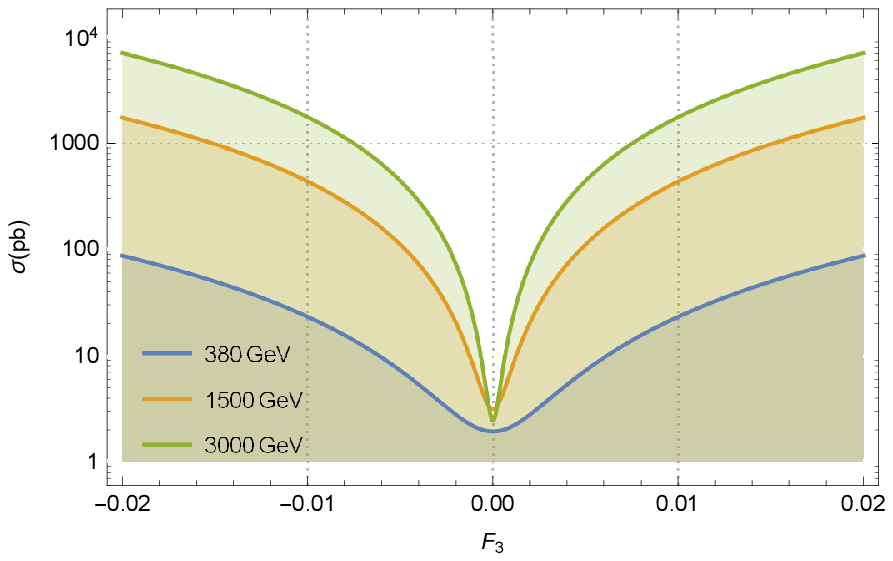}}}
\caption{ \label{fig:gamma2} Same as in Fig. 2, but for $F_3$.}
\end{figure}

\begin{figure}[t]
\centerline{\scalebox{0.8}{\includegraphics{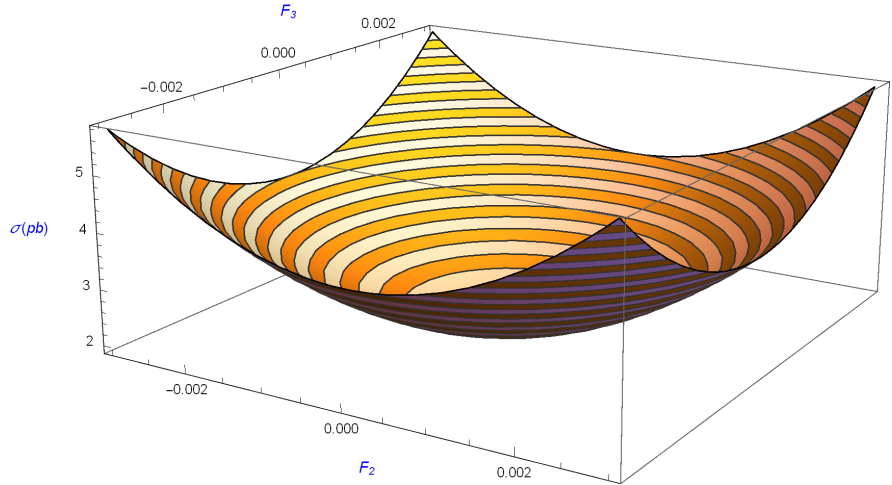}}}
\caption{ \label{fig:gamma1} The total cross-sections of the process
$\gamma e^- \to \tau \bar\nu_\tau \nu_e $ as a function of $F_2$ and $F_3$
for center-of-mass energy of $\sqrt{s}=380\hspace{0.8mm}GeV$.}
\end{figure}

\begin{figure}[t]
\centerline{\scalebox{0.8}{\includegraphics{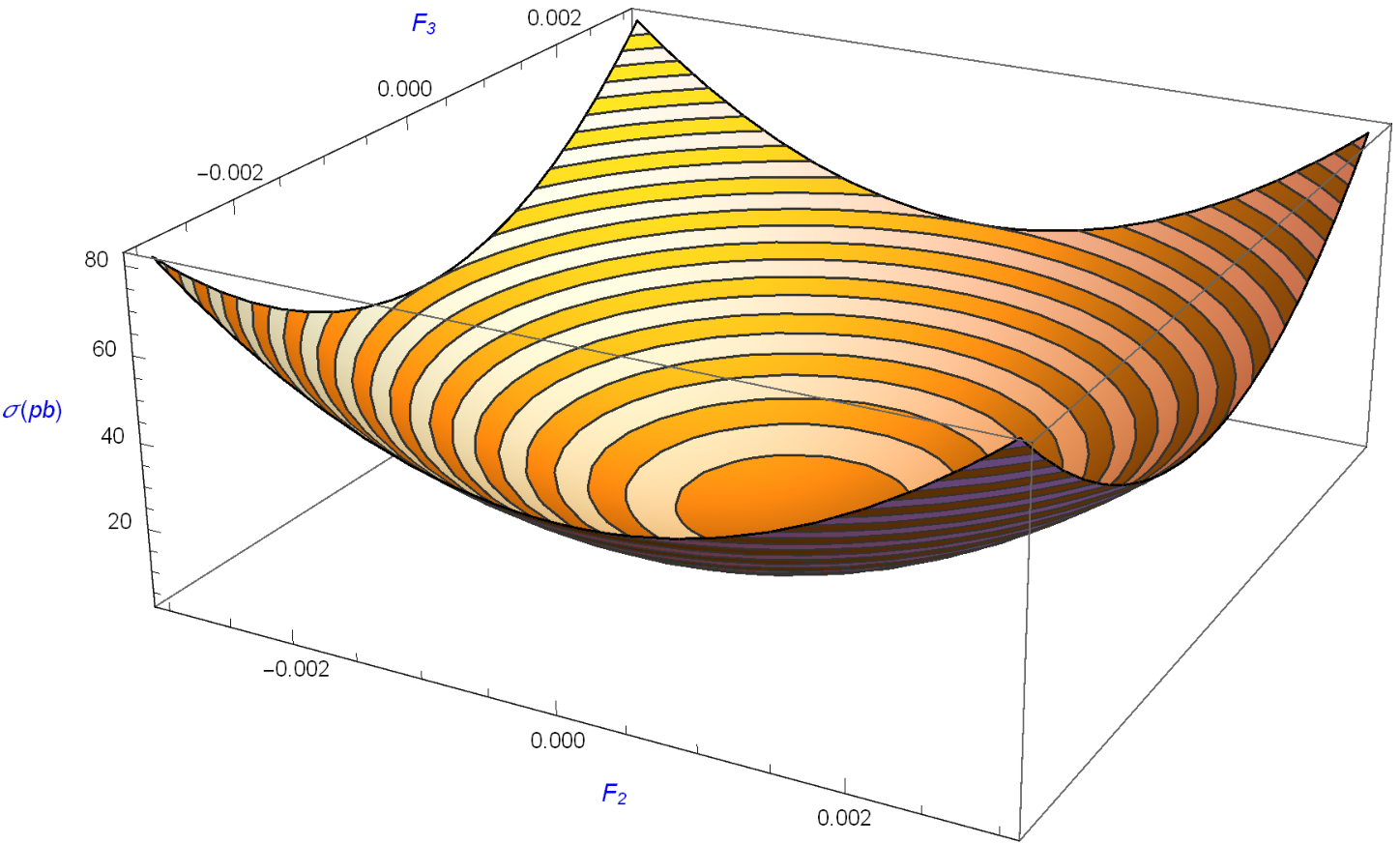}}}
\caption{ \label{fig:gamma3} Same as in Fig. 4, but for $\sqrt{s}=1500\hspace{0.8mm}GeV$.}
\end{figure}

\begin{figure}[t]
\centerline{\scalebox{0.8}{\includegraphics{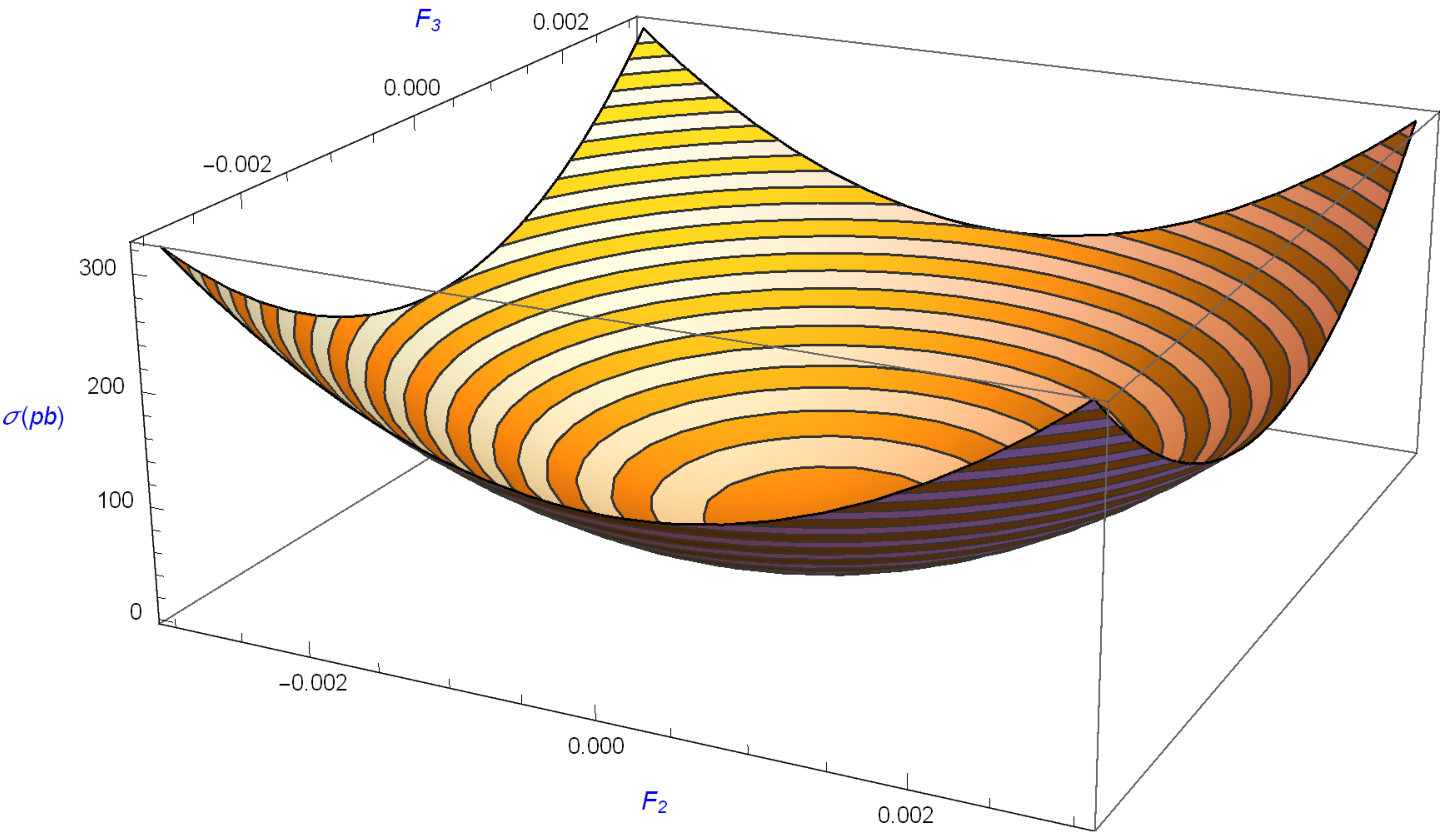}}}
\caption{ \label{fig:gamma4} Same as in Fig. 4, but for $\sqrt{s}=3000\hspace{0.8mm}GeV$.}
\end{figure}

\begin{figure}[t]
\centerline{\scalebox{1}{\includegraphics{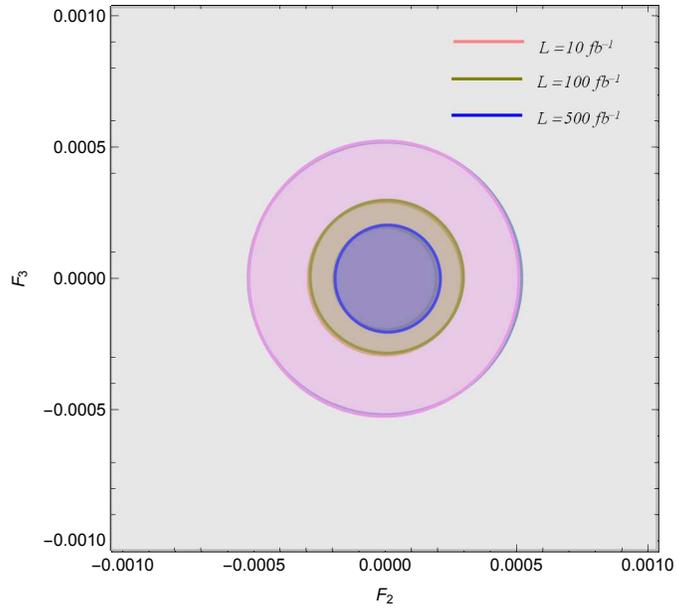}}}
\caption{ \label{fig:gamma2} Bounds contours at the $95\% \hspace{1mm}C.L.$ in the
$F_3-F_2$ plane for the process $\gamma e^- \to \tau \bar\nu_\tau \nu_e $ with the
$\delta _{sys}=0\%$ and for center-of-mass energy of $\sqrt{s}=380\hspace{0.8mm}GeV$.}
\end{figure}

\begin{figure}[t]
\centerline{\scalebox{1}{\includegraphics{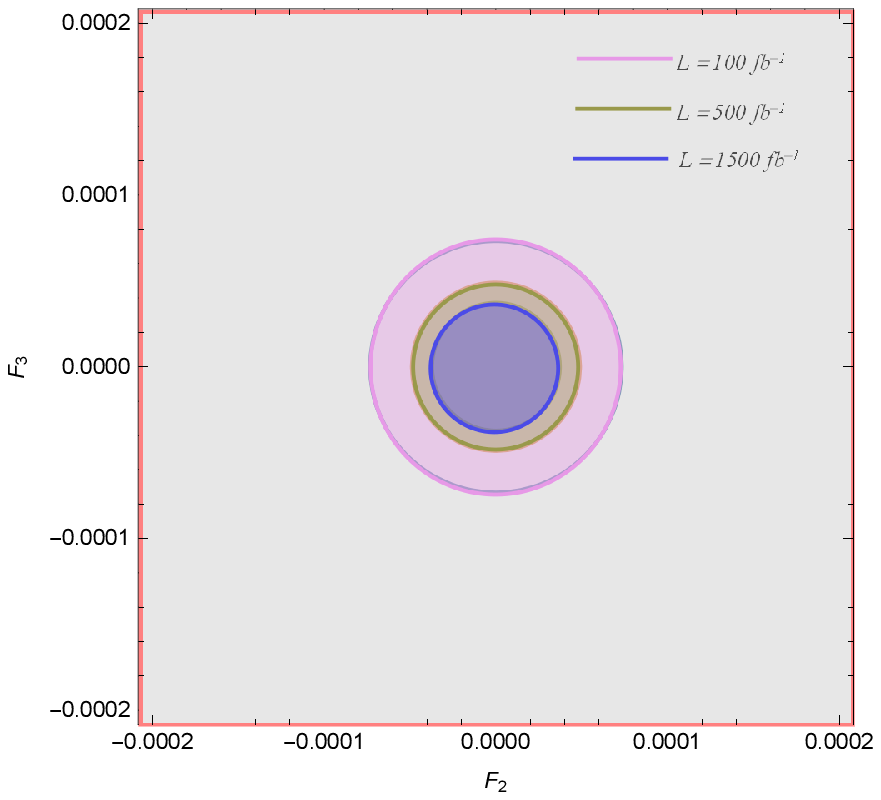}}}
\caption{ \label{fig:gamma4} Same as in Fig. 7, but for $\sqrt{s}=1500\hspace{0.8mm}GeV$.}
\end{figure}

\begin{figure}[t]
\centerline{\scalebox{1}{\includegraphics{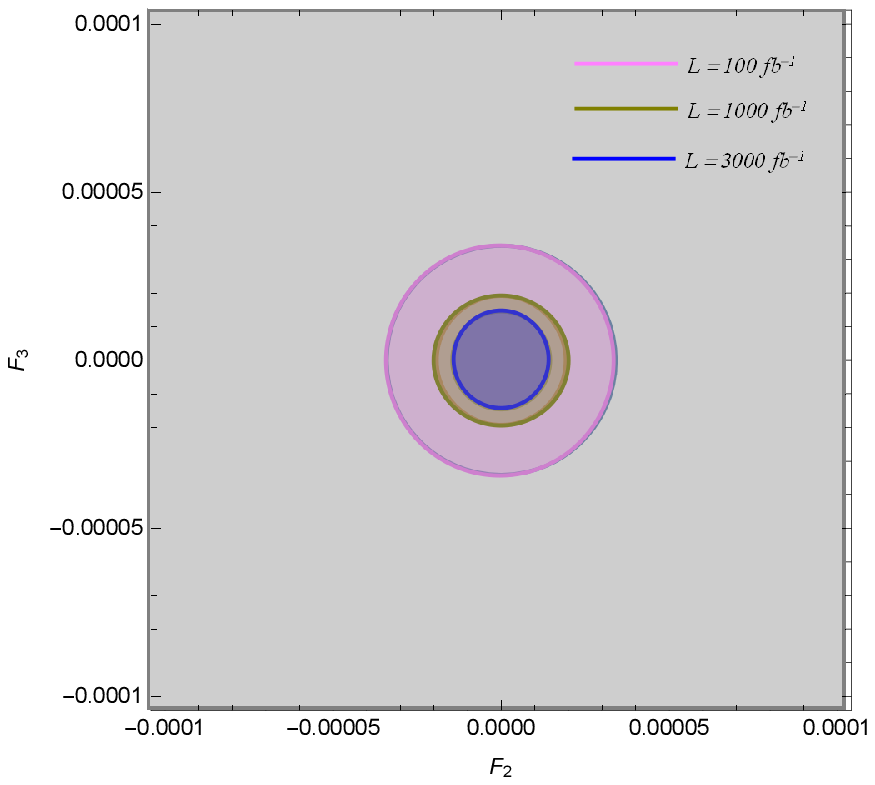}}}
\caption{ \label{fig:gamma4} Same as in Fig. 7, but for $\sqrt{s}=3000\hspace{0.8mm}GeV$.}
\end{figure}

\begin{figure}[t]
\centerline{\scalebox{1.2}{\includegraphics{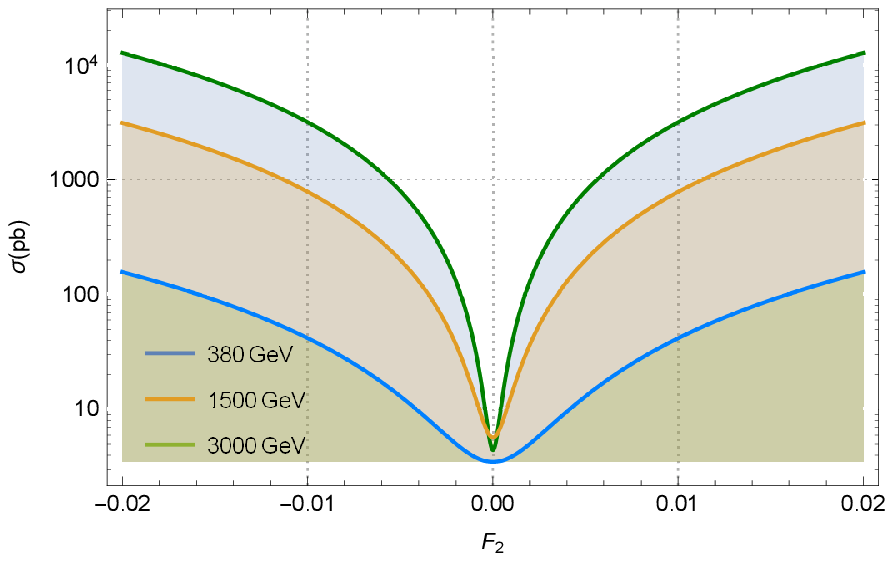}}}
\caption{ \label{fig:gamma4} Same as in Fig. 2, but with polarized electron beams $P_e=-80\%$.}
\end{figure}

\begin{figure}[t]
\centerline{\scalebox{1.2}{\includegraphics{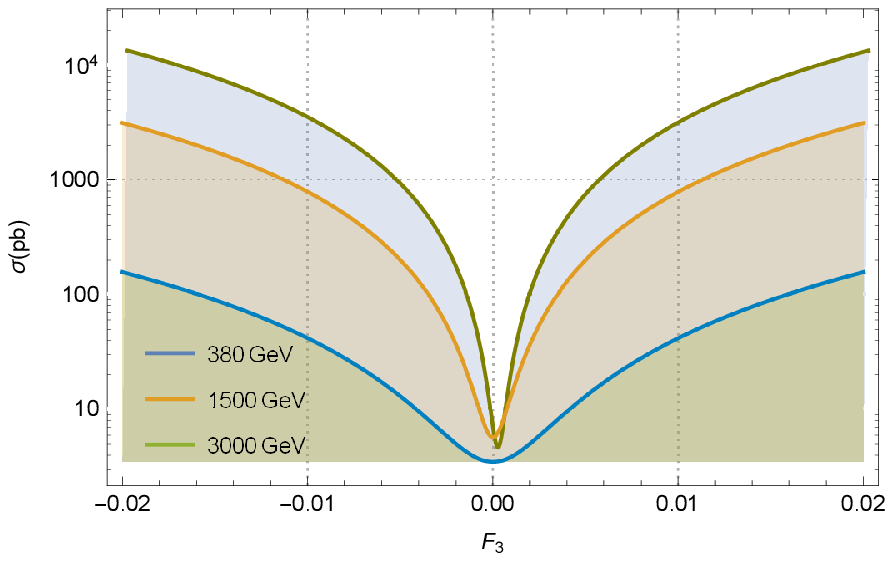}}}
\caption{ \label{fig:gamma4} Same as in Fig. 3, but with polarized electron beams $P_e=-80\%$.}
\end{figure}

\begin{figure}[t]
\centerline{\scalebox{1.2}{\includegraphics{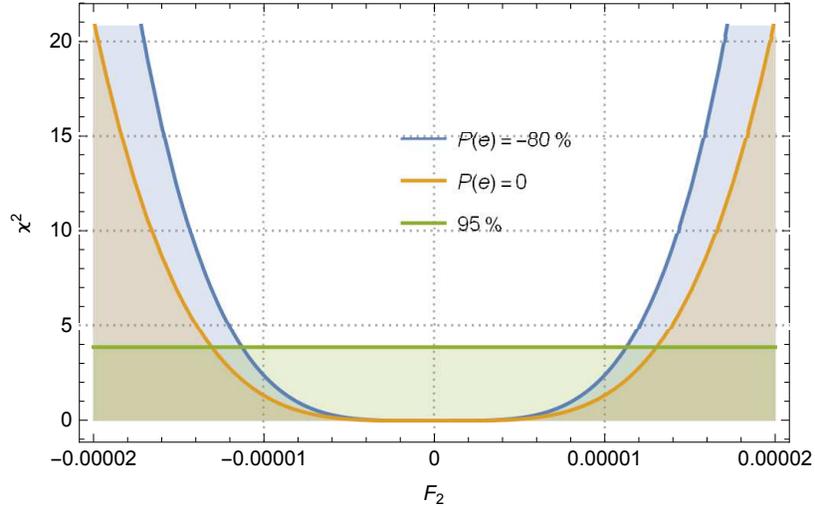}}}
\caption{ \label{fig:gamma4} $\chi^2$ as a function of $F_2$ for the total cross-section of the process
$\gamma e^- \to \tau \bar\nu_\tau \nu_e $.}
\end{figure}

\begin{figure}[t]
\centerline{\scalebox{1.2}{\includegraphics{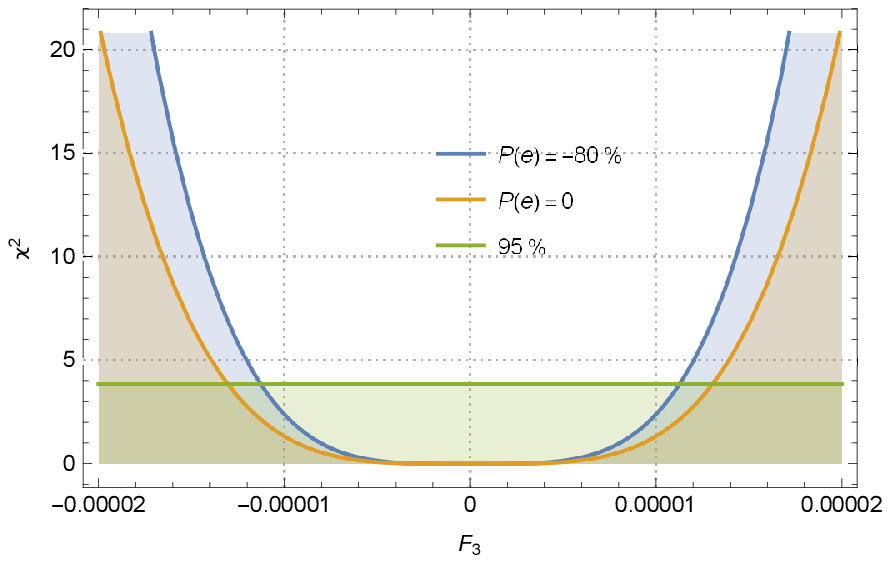}}}
\caption{ \label{fig:gamma4} Same as in Fig. 12, but for $F_3$.}
\end{figure}

\begin{figure}[t]
\centerline{\scalebox{1.2}{\includegraphics{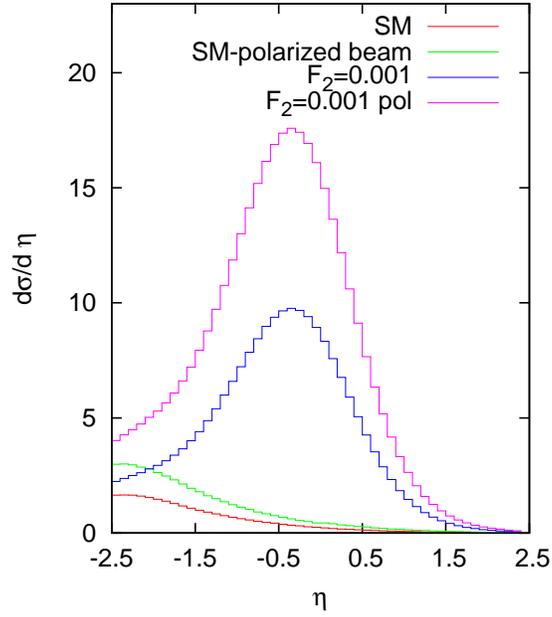}}}
\caption{ \label{fig:gamma4} Generated tau-lepton pseudorapidity distribution for
$\gamma e^- \to \tau \bar\nu_\tau \nu_e $. The distributions are for SM (SM-polarized beam)
and $F_2$ ($F_2$-polarized beam).}
\end{figure}

\begin{figure}[t]
\centerline{\scalebox{1.2}{\includegraphics{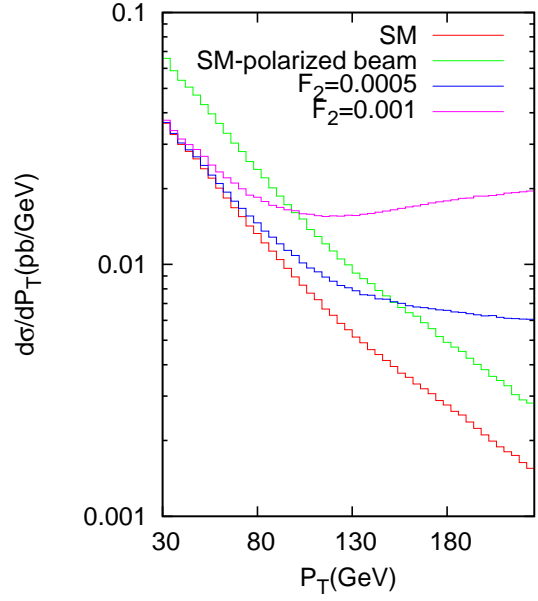}}}
\caption{ \label{fig:gamma4} Same as in Fig. 14, but for $P_T$.}
\end{figure}

\end{document}